\documentclass[sigconf]{acmart}
\usepackage{slnotation}
\usepackage{fdiaz}
\usepackage{subcaption}
\usepackage{booktabs} 
\usepackage{tcolorbox}
\usepackage{algpseudocode}
\usepackage{algorithm}
\usepackage{booktabs}
\usepackage{tabularx}
\usepackage[table]{xcolor}
\usepackage{makecell}
\usepackage{graphicx}
\usepackage{adjustbox}

\definecolor{genblue}{RGB}{100, 160, 210}    
\definecolor{retred}{RGB}{220, 160, 130}     
\definecolor{routergreen}{RGB}{120, 190, 150} 

\setcopyright{rightsretained}

\AtBeginDocument{%
  }

\copyrightyear{2026}
\acmYear{2026}
\setcopyright{cc}
\setcctype{by}

\acmConference[SIGIR '26] {Proceedings of the 49th International ACM SIGIR Conference on Research and Development in Information Retrieval}{July 20--24, 2026}{Melbourne, VIC, Australia.}
\acmBooktitle{Proceedings of the 49th International ACM SIGIR Conference on Research and Development in Information Retrieval (SIGIR '26), July 20--24, 2026, Melbourne, VIC, Australia}
\acmISBN{979-8-4007-2599-9/2026/07}
\acmDOI{10.1145/3805712.3808542}



\renewcommand{\slReals}{\mathbb{R}}

\renewcommand{\mlOutput}{\hat{y}}

\begin{document}
\title{Evaluation of Agents under Simulated AI Marketplace Dynamics}

\author{To Eun Kim}
\orcid{0000-0002-2807-1623}
\affiliation{%
  \institution{Carnegie Mellon University}
  \city{Pittsburgh}
  \state{PA}
  \country{United States}
}
\email{toeunk@cs.cmu.edu}

\author{Alireza Salemi}
\orcid{0009-0006-1937-2615}
\affiliation{%
  \institution{University of Massachusetts Amherst}
  \city{Amherst}
  \state{MA}
  \country{United States}
}
\email{asalemi@cs.umass.edu}

\author{Hamed Zamani}
\orcid{0000-0002-0800-3340}
\affiliation{%
  \institution{University of Massachusetts Amherst}
  \city{Amherst}
  \state{MA}
  \country{United States}
}
\email{zamani@cs.umass.edu}

\author{Fernando Diaz}
\orcid{0000-0003-2345-1288}
\affiliation{%
  \institution{Carnegie Mellon University}
  \city{Pittsburgh}
  \state{PA}
  \country{United States}
}
\email{diazf@acm.org}

\begin{abstract}
Modern information access ecosystems consist of mixtures of systems, such as retrieval systems and large language models, and increasingly rely on marketplaces to mediate access to models, tools, and data, making competition between systems inherent to deployment. In such settings, outcomes are shaped not only by benchmark quality but also by competitive pressure, including user switching, routing decisions, and operational constraints. Yet evaluation is still largely conducted on static benchmarks with accuracy-focused measures that assume systems operate in isolation. This mismatch makes it difficult to predict post-deployment success and obscures competitive effects such as early-adoption advantages and market dominance. We introduce \textit{Marketplace Evaluation}, a simulation-based paradigm that evaluates information access systems as participants in a competitive marketplace. By simulating repeated interactions and evolving user and agent preferences, the framework enables longitudinal evaluation and marketplace-level metrics, such as retention and market share, that complement and can extend beyond traditional accuracy-based metrics. We formalize the framework and outline a research agenda, motivated by business and economics, around marketplace simulation, metrics, optimization, and adoption in evaluation campaigns like TREC.
\end{abstract}
\maketitle
\section{Introduction}
\label{sec:introduction}

\begin{figure}[t]
  \centering
  \begin{subfigure}{0.32\linewidth}
    \includegraphics[width=\linewidth,
      trim=100 260 554 188,clip]{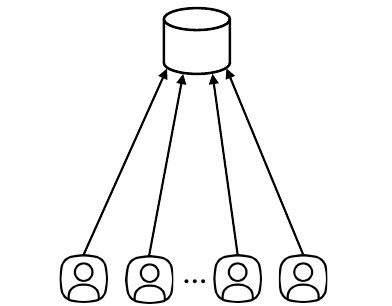}
    \caption{Cranfield}\label{fig:paradigms:multiuser}
  \end{subfigure}%
  \hfill
  \begin{subfigure}{0.32\linewidth}
  \hspace{-0.1\linewidth} 
    \includegraphics[width=\linewidth,
      trim=340 260 313 188,clip]{01-introduction/graphics/multifirm-framework-comparison.pdf}
    \caption{Arena}\label{fig:paradigms:arena}
  \end{subfigure}%
  \hfill
  \begin{subfigure}{0.32\linewidth}
    \includegraphics[width=\linewidth,
      trim=594 260 59 188,clip]{01-introduction/graphics/multifirm-framework-comparison.pdf}
\caption{Marketplace}\label{fig:paradigms:multifirm}
  \end{subfigure}
  \caption{
  Comparison of evaluation paradigms for information access systems. 
    (a)~\textbf{Cranfield}: a multi-user single-system setting where all queries are routed to a fixed system, eliminating competition and user choice. 
    (b)~\textbf{Arena}: a multi-user multi-system setting where a central platform mediates anonymous pairwise comparisons, precluding persistent user preferences or switching behavior. 
    (c)~\textbf{Marketplace}: a multi-user multi-system setting where users actively select among competing providers without requiring a central mediator, enabling competitive dynamics, behavioral adaptation, and the emergence of market share over time.
}
  \label{fig:evaluation_paradigms}
\end{figure}

Modern information access (IA) systems are increasingly interactive, often operating through conversational interfaces powered by large language models (LLMs) that deliver generated information objects (GIOs) to users \cite{swirl_18}. 
These systems, hereafter referred to as \textit{IA agents}, can be any system that provides information access functionality, including standalone LLMs, standalone retrieval systems, or compositions of one or more LLMs and retrievers \cite{zamani:reml}.

Development in this space has been rapid and highly competitive, with providers introducing new capabilities at a near-monthly pace \cite{korinek2025concentrating, longpre2025economiesOfOpenIntelligence}.
This competition plays out across marketplaces that mediate access to various information access agents, including open community hubs \cite{longpre2025economiesOfOpenIntelligence}, centralized execution platforms \cite{anthropic2024mcp, openrouter}, and enterprise marketplaces \cite{aws_marketplace, databricks_marketplace}. The existence of these platforms make it easy for users and systems to try multiple providers and switch when better options appear.

However, evaluation of IA agents is predominantly conducted using static benchmarks that treat systems as standalone artifacts \cite{wei_measuring_2024, trec-rag}. This isolated view is therefore misaligned with deployment: within marketplace, performance is inherently interdependent, since the success of one system depends not only on its intrinsic quality but also on the presence, behavior, and strategic choices of competing systems. These interactions fundamentally alter what it means for a system to be \textit{good}, shifting evaluation from absolute effectiveness to relative performance under competitive pressure.

Understanding system performance in the competitive marketplaces is crucial for both system designers and platform operators, as isolated benchmarks often fail to predict real-world adoption and success. For example, in retrieval-augmented generation (RAG) \cite{lewis:reml-nlg}, the interaction between generators and retrievers creates network effects where the value of a retrieval system depends not only on its intrinsic quality but also on its adoption by the generator population and its differentiation from competing systems \cite{porter1997competitive}. Marketplace dynamics introduce phenomena such as winner-take-all effects \cite{rosen1981economics}, where small performance differences can lead to dramatic market share disparities, and portfolio effects \cite{portfolio_selection}, where generators may strategically diversify across multiple retrieval systems to optimize for different query types or hedge against system failures. These dynamics have significant implications for research priorities, business strategies, and the overall health of information ecosystems, as they influence innovation incentives, market concentration, and the diversity of available approaches.

We present the perspective that existing evaluation paradigms in information access, including Cranfield-style evaluations \cite{cleverdon:cranfield-og} and pairwise A/B testing \cite{joachims:ranking-svm}, do not account for competitive market forces or user behavioral adaptation.
In the Cranfield paradigm (Figure \ref{fig:paradigms:multiuser}), evaluation can be interpreted as a multi-user single-system setting, where multiple queries stand in for multiple users but all interactions are routed to a fixed system, eliminating any notion of competition or user choice. 
Pairwise and Arena-style evaluations (Figure \ref{fig:paradigms:arena}) extend this to a multi-user multi-system setting, yet still fall short of modeling real-world competition, since users are typically presented with anonymous system variants and do not actively select among providers, preventing the formation of persistent preferences or switching behavior. 
As a result, the single-system evaluation fails to capture how system performance degrades or improves under varying load conditions as market share fluctuates, while pairwise comparisons cannot model the complex multi-way interactions that emerge when multiple systems compete simultaneously for the same user base. Furthermore, current approaches do not account for strategic behavior by system operators, such as quality differentiation that influence system selection beyond pure performance metrics. The temporal dynamics of marketplace competition, including learning effects, adaptation strategies, and the evolution of user preferences, are entirely absent from traditional evaluation frameworks, yet these factors critically determine long-term system viability and market outcomes.

Following this perspective, we introduce \textit{marketplace evaluation}, a general framework for assessing information access systems as participants in competitive marketplaces rather than as isolated artifacts. The framework models information access as a multi-agent system in which providers and intermediaries make strategic decisions under shared constraints, and system performance emerges from interactions among competing entities rather than from standalone effectiveness.

Without loss of generality, we ground this framework in the RAG ecosystem, a representative setting that encompasses all key interactive stakeholders, including users, generators, and retrievers. In this setting, for interactive information access, users choose among multiple competing generators, which differ in capability and cost, while generators in turn act as demand-side agents that select among competing retrieval services. Here, retrieval systems also compete for traffic coming from generators not only through retrieval quality, but also through capability, cost, and control over which collections of information they are able or permitted to serve. Both generators and retrievers may strategically restrict, expand, or specialize their accessible information sources, making access itself a central axis of competition.

Within this instantiation, we formalize evaluation with an agent-based simulation \cite{epstein1996growing, axtell2000whyagents, rand_agent-based_2011} of competitive information access marketplaces, in which system performance depends on adoption patterns, competitive pressure, and information access constraints. Our framework incorporates realistic marketplace factors such as capacity limits, pricing mechanisms, and quality-of-service guarantees, enabling systematic evaluation of how systems perform, adapt, and coexist under different competitive structures.

We begin in Section \ref{sec:motivation} with our perspective and a motivating experiment that illustrates how competitive dynamics alter system-level conclusions. We then formalize the marketplace evaluation framework with consistent notations in Section \ref{sec:framework}.
Building on this foundation, we outline a long term research program centered on three core research questions:
\begin{itemize}[leftmargin=1em]
    \item RQ1 (\S\ref{sec:rq1}): How can information access agent marketplaces be simulated for evaluation?
    \item RQ2 (\S\ref{sec:rq2}): What metrics are appropriate for characterizing IA agent's performance and market conditions?
    \item RQ3 (\S\ref{sec:rq3}): How can marketplace evaluation be integrated into existing evaluation campaigns and benchmarking infrastructures?
\end{itemize}

We hope this fresh perspective inspires a new class of methodologies for evaluating and improving the impact of information retrieval technologies on AI ecosystems, while addressing one of the most pressing challenges in modern AI.

\section{Perspective: Evaluation of IA Agents as Marketplace Participants}\label{sec:motivation}

We argue that IA agents should be evaluated as competitors in a marketplace, where performance emerges through repeated interaction and both human and agent preferences evolve over time.
In such environments, systems compete for attention, usage, and continued engagement from users.
Consequently, evaluation outcomes depend on how systems are exposed to users, how preferences form over time, and how competition shapes long-term usage patterns.

Current evaluation methods are fundamentally limited by the assumption that interactions are independent and isolated, failing to capture how humans and AI agents evolve through experience. Traditional static benchmarks and Cranfield paradigm ignore the fact that human preferences are shaped by prior exposure. In the real world, the specific sequence and timing of interactions build the loyalty or abandonment that dictates market success \cite{exposure_effect}. Similarly, modern multi-agent IA  systems often utilize \textit{routers} \cite{ong2025routellm, tang2025adapting, jeong-etal-2024-adaptive, kim2025ltrr} to dynamically direct queries to the most appropriate retriever or generator. These components also develop their own internal tendencies over time by learning from feedback such as cost, latency, and accuracy. Consequently, they exhibit \textit{path dependence} where early interactions alter future system behavior and resource allocation. By treating every interaction as a first time occurrence, current paradigms miss the cumulative, historical dynamics that define real-world performance.
This misalignment motivates a rethinking of evaluation paradigms toward settings and metrics that explicitly model and evaluate interaction, adaptation, and competition.

This evaluation perspective algins with common business and deployment goals. 
For industry practitioners, the success of an IA agents is rarely defined by a single offline metric \cite{netflix_recsys, jannach2019measuring}.
Instead, deployment success is reflected through operational signals such as query traffic, market share, and sustained usage \cite{rodden2010measuring}.
However, this perspective does not diminish the importance of accuracy-based metrics (e.g., NDCG \cite{wang2013theoretical} for ranking evaluation, and ROUGE \cite{lin2004rouge} for GIO evaluation), as those quality do influence user satisfaction and underlies downstream outcomes such as retention and growth \cite{dan2016measuring}.
Marketplace-level metrics therefore complement the traditional metrics, rather than replacing them.

\subsection{Motivating Experiment}\label{sec:motivating-exp}
To illustrate what evaluation under competition reveals that traditional static benchmarks cannot and to motivate a future research agenda, we evaluate multiple IA agents using a marketplace simulation with a simple setup.

We first conduct static evaluations to create a ranking of systems.
We set up seven distinct systems: 
DeepSeek V3.2 \cite{liu2025deepseek}, Kimi K2.5 \cite{team2026kimi}, Gemini 2.5 \cite{comanici2025gemini}, GPT-OSS \cite{agarwal2025gptoss}, Grok 4.1 \cite{xai_grok}, Qwen3 \cite{yang2025qwen3}, and Llama 3.3 \cite{dubey2024llama}. These agents are tested using 500 fact-seeking questions drawn from the test set of the SimpleQA benchmark. \cite{wei_measuring_2024}. 
These agents produce generated information objects (GIO) \cite{swirl_18} instead of being restricted to short-phrase responses.
Each agent answers all questions, responses are scored using a question-level correctness metric, and systems are ranked by the aggregated metric (F-score), following the benchmark.
This procedure yields a single, static system ranking that is invariant to evaluation order and independent of user interaction. The evaluation results and the system ranking can be found in Table \ref{tab:static_results}.

\subsubsection{\textbf{Marketplace Simulation Setup.}}
We contrast this static evaluation with a minimal marketplace simulation that preserves the same questions and the same correctness metric, but introduces user choice and preference updating.
We simulate a population of 10 users who can access all systems in the market and maintain individual preference distributions over systems.
Per simulation step, 5 users are sampled, and each user is assigned to a question drawn without replacement from the dataset pool, and the user selects a system based on their current preference distribution with a small amount of exploration. 
After observing the system response, the user updates their preference based solely on the question-level correctness of that response.

\begin{figure*}[t]
    \vspace{-0.4cm}
    \centering
    \begin{subfigure}{0.49\textwidth}
        \centering
        \includegraphics[width=\linewidth, trim=0 0 0 0 , clip]{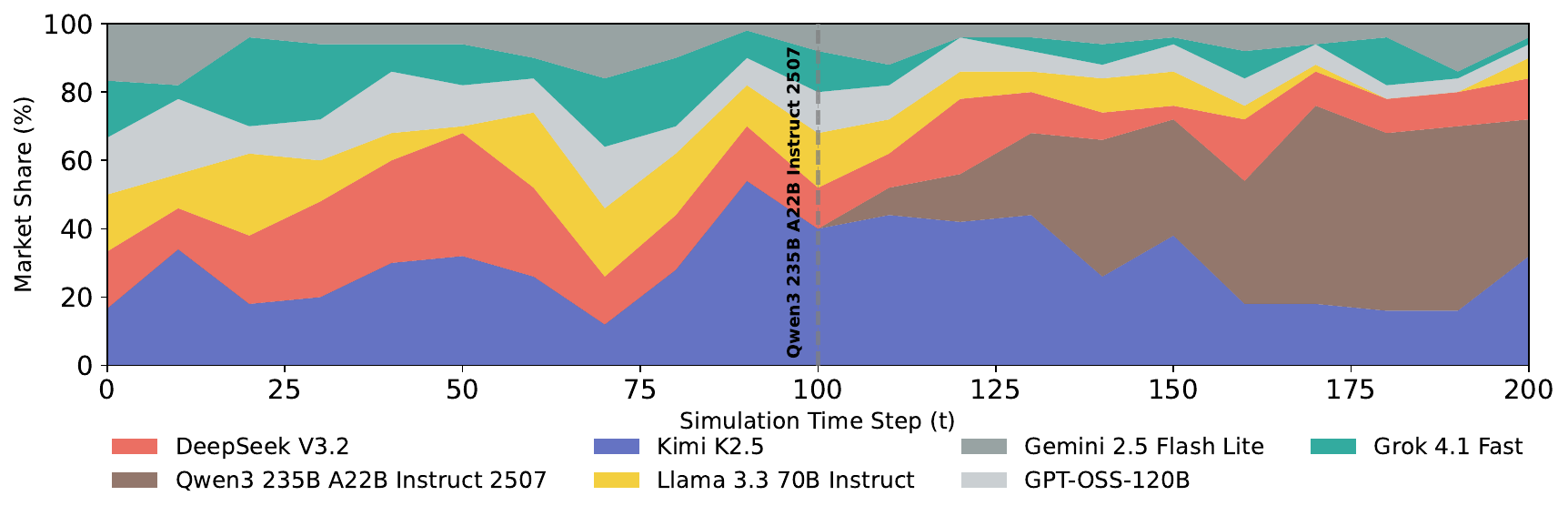}
        \vspace{-0.4cm}
        \caption{Qwen3-235B-A22B-Instruct-2507 introduced at $t=100$.\\Late entry of a strong model into a lightly concentrated market.}
        \label{fig:marketshare-exp:qwen-later}
    \end{subfigure}
    \hfill
    \begin{subfigure}{0.49\textwidth}
        \centering
        \includegraphics[width=\linewidth, trim=0 0 0 0, clip]{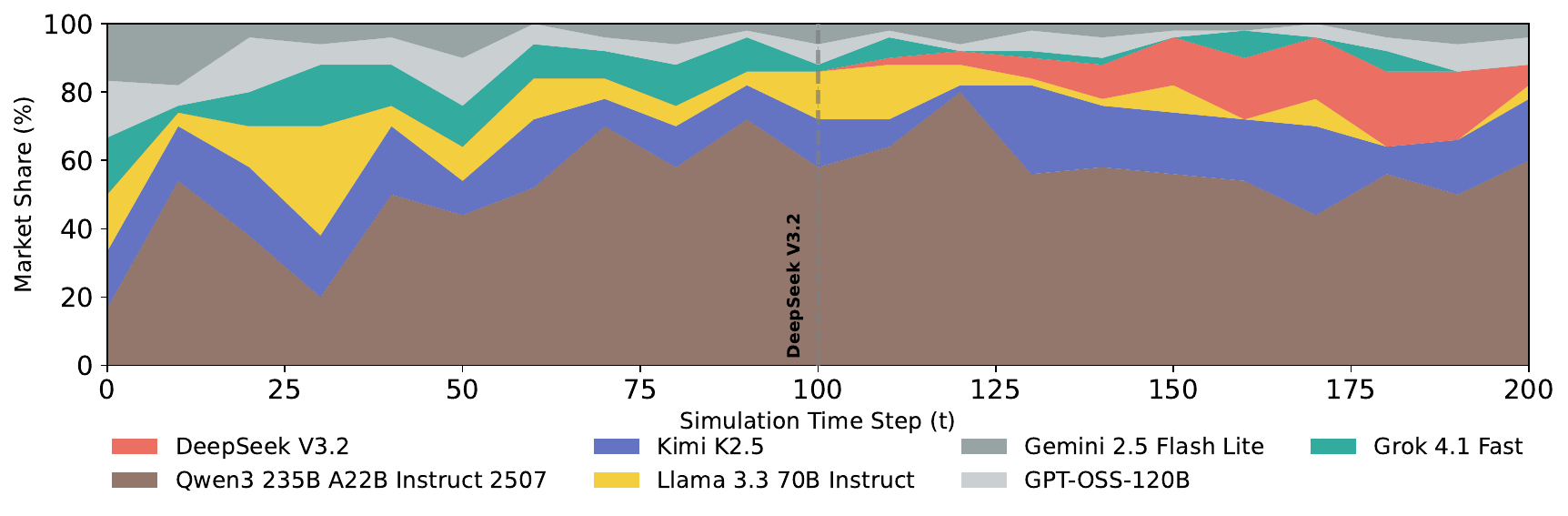}
        \caption{DeepSeek V3.2 introduced at $t=100$.\\Late entry of an average model into a highly concentrated market.}
        \label{fig:marketshare-exp:ds-later}
    \end{subfigure}
    \vspace{-0.2cm}
    \caption{Marketplace dynamics illustrating fluctuating system rankings and market share following the entry of a new model. The first 100 simulation steps were conducted with six models, after which a seventh model was introduced into the warm started marketplace.}
    \label{fig:marketshare-exp}
\end{figure*}

This way, the first 100 simulation steps cover the same 500 questions used in our static evaluation. We initially run these steps with six models, introducing a seventh model during the second half of the simulation (steps 101 to 200). This setup enables us to examine two contrasting scenarios under an active market with warm-started users. In the first, a system that performs strongly under static benchmarking, Qwen3, enters a lightly concentrated market (Table \ref{tab:qwen_later} \& Figure~\ref{fig:marketshare-exp:qwen-later}). In the second, a system with moderate static benchmark performance, DeepSeek V3.2, enters a highly concentrated market (Table \ref{tab:ds_later} \& Figure~\ref{fig:marketshare-exp:ds-later}).
%

\begin{table}[!t]
\centering
\caption{Model performance (F1-score) and fair market shares proportionate to the benchmark performance.}
\vspace{-0.4cm}
\label{tab:model_fairshare}
\adjustbox{max width=\linewidth}{
\begin{tabular}{lcccc}
\hline
\textbf{Model} & \textbf{F1} & \textbf{Fair Share (\%)} & \makecell{\textbf{Fair Share (\%)} \\ \textbf{w/o Qwen3}}  & \makecell{\textbf{Fair Share (\%)} \\ \textbf{w/o DeepSeek V 3.2}} \\
\hline
Qwen 3        & 60.24 & 28.89 & - & 33.30 \\
Kimi K2.5     & 42.51 & 20.38 & 28.66 & 23.49 \\
Llama 3.3     & 27.74 & 13.30 & 18.71 & 15.33 \\
DeepSeek V3.2 & 27.59 & 13.23 & 18.60 & - \\
Grok 4.1      & 19.21 & 9.21 & 12.95 & 10.62 \\
Gemini 2.5    & 16.83 & 8.07 & 11.35 & 9.30 \\
GPT-OSS       & 14.42 & 6.91 & 9.72 & 7.97 \\
\hline
\end{tabular}}
\vspace{-10pt}
\label{tab:static_results}
\end{table}


\begin{table*}[t]
    \centering
    \begin{minipage}{0.49\textwidth}
        \centering
        \caption{\textcolor{brown}{Qwen3} late entry.}
        \vspace{-0.4cm}
        \footnotesize 
        \begin{tabular*}{\textwidth}{@{\extracolsep{\fill}}cc c cc@{}}
        \hline
        \makecell{Market Share\\($\Delta$FS) (\%)}  & \makecell{System\\Ranking} & $\rightarrow$ & \makecell{System\\Ranking} & \makecell{Market Share\\(\%) ($\Delta$FS)} \\ 
        \hline
        \multicolumn{2}{c}{\makecell{Before Entry\\HHI = 1940.96\\$t = (1 - 100)$}} &  & \multicolumn{2}{c}{\makecell{After Entry\\HHI = 2430.16\\$t = (101 - 200)$}} \\ \hline
             &                & new & \cellcolor{brown!70}Qwen3         & 36.00 (\textcolor{green!60!black}{+7.11}) \\
        (\textcolor{green!60!black}{+0.74}) 29.40 & Kimi K2.5     &  & Kimi K2.5    & 29.40 (\textcolor{green!60!black}{+8.62}) \\ 
        (\textcolor{green!60!black}{+2.4}) 21.00 & DeepSeek V3.2 &  & DeepSeek V3.2 & 11.60 (\textcolor{red}{-1.63})\\
        (\textcolor{red}{-4.31}) 14.40 & Llama 3.3      &  & Gemini 2.5    & 6.60 (\textcolor{red}{-1.47}) \\ 
        (\textcolor{green!60!black}{+0.85}) 13.80 & Grok 4.1       &  & GPT-OSS       & 6.40 (\textcolor{red}{-0.51}) \\
        (\textcolor{green!60!black}{+3.08}) 12.80 & GPT-OSS        &  & Llama 3.3     & 5.60 (\textcolor{red}{-7.7}) \\ 
        (\textcolor{red}{-2.75}) 8.60  & Gemini 2.5     &  & Grok 4.1      & 4.40 (\textcolor{red}{-4.81}) \\
        \hline
        \end{tabular*}
        \label{tab:qwen_later}
    \end{minipage}
    \hspace{0.1cm} 
    \begin{minipage}{0.49\textwidth}
        \centering
        \caption{\textcolor{red!70}{DeepSeek V3.2} late entry.}
        \vspace{-0.4cm}
        \footnotesize 
        \begin{tabular*}{\textwidth}{@{\extracolsep{\fill}}cc c cc@{}}
        \hline
        \makecell{Market Share\\($\Delta$FS) (\%)}  & \makecell{System\\Ranking} & $\rightarrow$ & \makecell{System\\Ranking} & \makecell{Market Share\\(\%) ($\Delta$FS)} \\ 
        \hline
        \multicolumn{2}{c}{\makecell{Before Entry\\HHI = 3176.88\\$t = (1 - 100)$}} &  & \multicolumn{2}{c}{\makecell{After Entry\\HHI = 3789.28\\$t = (101 - 200)$}} \\ \hline
        (\textcolor{green!60!black}{+22.71}) 51.60 & Qwen3     &  & Qwen3    & 57.80 (\textcolor{green!60!black}{+28.91}) \\ 
        (\textcolor{red}{-5.58}) 14.80 & Kimi K2.5 &  & Kimi K2.5 & 15.80 (\textcolor{red}{-4.58})\\
                                &  & new & \cellcolor{red!30}DeepSeek V3.2 & 12.0 (\textcolor{red}{-1.23})\\
        (\textcolor{red}{-2.70}) 10.60 & Llama 3.3      &  & Llama 3.3    & 4.60 (\textcolor{red}{-8.70}) \\ 
        (\textcolor{green!60!black}{+0.39}) 9.60 & Grok 4.1       &  & GPT-OSS       & 4.20 (\textcolor{red}{-2.71}) \\
        (\textcolor{green!60!black}{+0.49}) 7.40 & GPT-OSS        &  & Gemini 2.5     & 3.20 (\textcolor{red}{-4.87}) \\ 
        (\textcolor{red}{-2.07}) 6.00  & Gemini 2.5     &  & Grok 4.1      & 2.40 (\textcolor{red}{-6.81}) \\
        \hline
        \end{tabular*}
        \label{tab:ds_later}
    \end{minipage}
\end{table*}

\subsubsection{\textbf{Ranking Divergence Under Interaction.}}
Let us begin by comparing the system rankings derived from the static benchmark (Table~\ref{tab:static_results}) with those obtained from the first set of simulation runs prior to the entry of any new models ($t=1\text{--}100$) in Tables~\ref{tab:qwen_later} and \ref{tab:ds_later}. Although the same set of questions was used, the resulting rankings differ, indicating that competitive dynamics and user adaptation reshape outcomes beyond what is captured by static evaluation alone. This discrepancy serves as one indicator that marketplace interactions can meaningfully alter system standing even before any structural changes, such as new entry, occur.

This becomes even clearer in examining the market share trends in Figures~\ref{fig:marketshare-exp:qwen-later} and \ref{fig:marketshare-exp:ds-later}. Market share is defined as the total query traffic received by an IA agent within a given time window. The plots report windowed market share with a size of 10, allowing us to track local trends as time progresses.
When systems are ranked by cumulative market share over $t=1\text{--}100$, a single overall ordering emerges. However, the windowed curves reveal that local rankings fluctuate frequently. In several periods, the market share of a model drops close to zero before recovering, or vice versa. These fluctuations highlight the underlying dynamics of user adaptation and competition that are not visible in aggregate statistics alone.

\subsubsection{\textbf{Market Entry and Dominance Effects.}}

More interesting effects emerge when the marketplace participants changes. From the windowed market share plots at $t=100$, the market in Figure~\ref{fig:marketshare-exp:qwen-later} is visibly less concentrated than in Figure~\ref{fig:marketshare-exp:ds-later}. The entry of Qwen3 acts as a clear shock to the market, whereas the entry of DeepSeek V3.2 produces only modest shifts, suggesting that entering an already highly concentrated market makes it difficult for a new model to secure the share implied by its standalone performance.

From a meritocratic perspective, one might expect market share to be proportional to internal capability, as approximated by static benchmark performance. Table~\ref{tab:static_results} reports the models' ``fair" shares derived from their F1 scores. However, when we compare these expected fair shares (FS) with the realized market shares, the discrepancy is substantial ($\Delta$FS in Tables \ref{tab:qwen_later} and \ref{tab:ds_later}). In particular, post entry market share disparity is notably higher than pre entry disparity, as reflected by the HHI index reported in Tables~\ref{tab:qwen_later} and \ref{tab:ds_later}.

In Figure~\ref{fig:marketshare-exp:qwen-later}, the entry of Qwen3 compresses the market shares of middle and lower ranked models into nearly indistinguishable levels. At that stage, the ranking implied by static benchmarking becomes largely uninformative. Across both scenarios, we observe dominance effects that resemble real world marketplaces and cannot be inferred from static benchmarking alone. The marketplace metrics discussed here, including market share and HHI, are examined in greater detail in Section~\ref{sec:rq2}.

\section{The Marketplace Evaluation Framework}
\label{sec:framework}
Before we delve into the core research questions, in this section, we formalize the marketplace evaluation framework, including how interaction dynamics unfold across successive simulation steps. 

\subsection{Overview}
\label{sec:overview}
\begin{figure}[t]
    \centering
    \resizebox{0.9\columnwidth}{!}{ 
        \includegraphics[trim=250 110 230 220, clip]{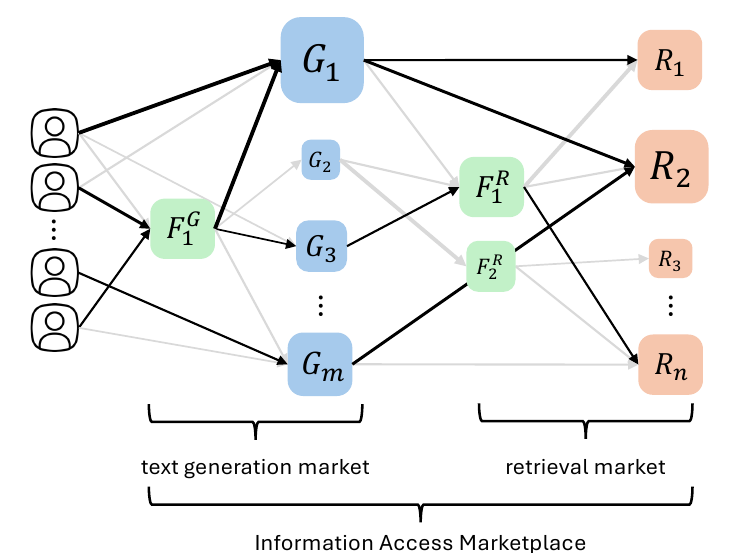}
    }
    \caption{An example snapshot of a RAG marketplace simulation. Nodes represent agents grouped by stakeholder role: users, \textcolor{genblue}{generators}, \textcolor{retred}{retrievers}, and \textcolor{routergreen}{routers} to generators and retrievers. Arrow thickness reflects selection preference intensity, node size reflects accumulated market share, and arrow opacity indicates whether a connection was selected or \textcolor{gray}{not selected} in the current snapshot.}
    \label{fig:simulation-snapshot}
    \vspace{-13pt}
\end{figure}

To study and evaluate systems within complex ecosystems, we formalize our framework using agent-based simulation \cite{epstein1996growing, axtell2000whyagents, rand_agent-based_2011}.

The dynamics observed in competitive information access agent marketplaces arise from interactions among many decision-makers, including human users and agents, coupled through feedback, competition, and path dependence.
For such systems, analytical characterization quickly becomes intractable and/or insoluble.
As argued by \citet{axtell2000whyagents} and, in the context of business research by \citet{rand_agent-based_2011}, agent-based simulation is not merely a modeling convenience but a necessary methodology for studying complex systems in which aggregate behavior emerges from interactions among heterogeneous agents.

Moreover, simulation provides a controlled yet expressive foundation for evaluation \cite{balog_user_2024, xu2025fairness_sigir25_tutorial}.
It enables the study of both short- and long-term deployment effects, supports scalable and reproducible experimentation without reliance on live user studies, and allows situated and counterfactual analysis under specific user populations and competitive conditions \cite{yao_-bench_2024, huang_crmarena-pro_2025}.
Recent advances in large language models make such simulations increasingly feasible \cite{filippas_large_2024, karten_llm_2025, park:generative-agents, zhang_simulating_2025}, positioning simulation-based evaluation as a practical mean to measure agents in the real-world marketplaces.

Figure~\ref{fig:simulation-snapshot} illustrates a concrete example of the agent-based simulation setting, involving users, generators, retrievers, and routers.
Consider a domain expert and a general web user interacting with three generator options: a high-capability general model, a moderate general-purpose model, and a domain-specialized model. The generators can access two retrieval back-ends: a public web search retriever and a proprietary domain retriever such as a legal document index. A router mediates requests from generators to retrievers based on routing policies. Generators may either route queries through the router or directly call a retriever, creating heterogeneous interaction patterns that evolve over time.

In this RAG ecosystem simulation example \cite{zamani:reml}, a single interaction unfolds as follows:
(i) a user issues a request and either directly selects, or is routed to, a generator;
(ii) the selected generator may in turn select a retriever to obtain supporting evidence before producing a response;
(iii) the resulting output is evaluated to produce a utility signal, which is then used to update future selection behavior of both users and generators.
As this process repeats across many interactions and many users, traffic allocation, agent preferences, and market share evolve over time.

Marketplace evaluation focuses on these \emph{longitudinal outcomes} with an ecosystem-centric view \cite{xu2025fairness_sigir25_tutorial}.
Rather than scoring agents in isolation on one-shot effectiveness, we evaluate how agents attract demand, retain usage, and remain viable under competition.
The remainder of this section formalizes the components and dynamics underlying such simulations.

Before turning to our core research agenda, we first formalize the agent-based marketplace simulation used for dynamic evaluation.

\subsection{Marketplace Primitives}\label{subsec:framework-primitives}

\paragraph{\textbf{Stakeholders.}}
We define a \emph{stakeholder} as a population of agents sharing a common \emph{functional role} in the marketplace. 
Grounding this definition in the example from Section~\ref{sec:overview}, the users, generators, retrievers, and router each constitute distinct stakeholder populations within the same marketplace. 
Roles here are defined by the task an agent performs, rather than by its internal architecture.
For example, the same language model may act as a generator when producing responses, or as a retriever when ranking documents.
Let $\mathcal{A} = \{\mathcal{U}, \mathcal{G}, \mathcal{R}, \mathcal{F}, \ldots\}$
denote the universe of possible stakeholder populations (users, generators, retrievers, routers, etc.).
A concrete evaluation instantiates a finite subset $A \subseteq \mathcal{A}$.
Each stakeholder $s \in A$ contains a set of agents with internal state parameters $\{\theta_{s,a}\}_{a \in s}$.
We describe common stakeholders and representative algorithmic choices from the literature in Section \ref{sec:rq1}.

\paragraph{\textbf{Markets and the Information Access Marketplace.}}
Competition arises whenever multiple agents provide functionally substitutable services for the same task and therefore compete for traffic. 
In the example from Section~\ref{sec:overview}, competition for user traffic arises not only among generators and among retrievers, but also across roles, as both generators and retrievers seek to capture and retain demand within the marketplace.
We define a \emph{market} as a task-specific group of stakeholders that jointly realize a functionality and compete for exposure within that functionality.
An \emph{information access marketplace} consists of one or more such markets.
Prominent examples include a \emph{text generation market} (generators competing to respond to user queries) and a \emph{document retrieval market} (retrievers competing to supply evidence), as depicted in Figure \ref{fig:simulation-snapshot}.

Between different markets, collaboration can arise through compositional workflows.
Retrieval-enhanced generation \cite{zamani:reml} illustrates how the retrieval market and the text generation market collaborate to improve end-user outcomes by incorporating retrieved evidence into generation.
At the same time, competition \emph{between} markets can still emerge when users choose among different information access modalities (e.g., selecting a conversational search versus a web search).

\paragraph{\textbf{Marketplace Governance Graph.}}
Inspired by \citet{zhuge_gptswarm_2024} and \citet{hoveyda_adaptive_2025}, the interaction structure of an instantiated marketplace (Figure \ref{fig:simulation-snapshot}) is defined by a directed acyclic graph
\begin{equation}
    G_{\text{marketplace}} = (A, \mathcal{E}),
\end{equation}
where nodes correspond to instantiated stakeholders in $A$.
An edge $(x \rightarrow y) \in \mathcal{E}$ indicates an \emph{admissible selection (invocation)} relation: an agent in stakeholder $x$ may select (i.e., call, route to, or delegate to) an agent in stakeholder $y$ as part of its computation.
Equivalently, the governance graph can be represented by an adjacency matrix
\begin{equation}
    W \in \{0,1\}^{|A|\times |A|},
\end{equation}
where $W[x,y]=1$ iff $(x \rightarrow y)\in\mathcal{E}$, otherwise $W[x,y]=0$.
Importantly, $G_{\text{marketplace}}$ is a \emph{marketplace governance graph} specifying which selections are allowed at the level of stakeholder roles, rather than the (possibly iterative) execution trace within a single episode.

\paragraph{\textbf{Interaction Semantics and Dynamics.}}
We now formalize how interactions unfold within an instantiated marketplace.
Given a fixed marketplace governance graph and stakeholder populations, interaction semantics specify how agents are selected, how artifacts are produced, and how realized outcomes induce feedback and adaptation.
This formulation abstracts over architectural details and captures execution at the level of \emph{who interacts with whom}, rather than \emph{how} each agent internally computes its output.

Each stakeholder population $s \in A$ is associated with a stochastic policy $\Pr_s(\cdot \mid \mathrm{pa}(s); \theta_s)$, where $\mathrm{pa}(s) = \{ s' \in A \mid (s' \rightarrow s) \in \mathcal{E} \}$ denotes the set of parent stakeholders of $s$ in $G_{\text{marketplace}}$, and $\theta_s$ denotes the parameters of agents within $s$.
A single interaction instance is generated by sampling along the governance graph in topological order:
\begin{equation}
z \sim \prod_{s \in \mathrm{topo}(G_{\text{marketplace}})}
\Pr_s\!\left(\cdot \mid \mathrm{pa}(s); \theta_s\right),
\end{equation}
where $z$ denotes the realized trajectory of selected agents.

Given an interaction outcome $z$, evaluation produces a utility:
\begin{equation}
\mu \sim \Pr(\cdot \mid z, \mathcal{D}; \theta_\mu),
\end{equation}
where $\mathcal{D}$ denotes the evaluation dataset and $\theta_\mu$ denotes the evaluator model. Depending on the dataset, evaluation may be reference-based (e.g., relevance or accuracy \cite{wei_measuring_2024}) or reference-free (e.g., LLM-based judges \cite{es2024ragas}), depending on the application.

After observing $\mu$, participants may update their internal states:
\begin{equation}
\theta_{s^*} \leftarrow \textsc{Update}(s^*, z, \mu),
\end{equation}
where $s^*$ denotes any agent selected during $z$. Referring to the example in Section~\ref{sec:overview}, suppose the expert user selects the moderate general-purpose generator, which in turn relies on the general web retriever. If the returned result is unsatisfactory, the user may reduce its preference for this generator in expert-level tasks. This negative feedback can propagate upstream, discouraging the generator from selecting the same retriever in similar contexts. Conversely, positive outcomes can reinforce the current routing and selection behavior across all involved agents. Repeated interaction, evaluation, and adaptation induces a stochastic dynamical system over the instantiated stakeholders and agents.

\section{RQ1: Marketplace Simulation}\label{sec:rq1}
\begin{table*}[t]
\centering
\small
\caption{Prominent stakeholders in information access agent marketplaces, defined by functional roles. Each stakeholder adapts its behavior based on feedback from the simulated marketplace, and evaluation is conducted using longitudinal and market-level signals that complement traditional effectiveness metrics.}
\vspace{-0.3cm}
\renewcommand{\arraystretch}{1.4}
\begin{tabularx}{\textwidth}{l|X|X|X}
\hline
\rowcolor[gray]{0.85}
\textbf{Stakeholder} 
& \textbf{Goal} 
& \textbf{Adaptive Behavior in the Marketplace} 
& \textbf{Evaluation Focus} \\
\hline

Users 
& Sustained utility from the marketplace over time 
& Adapt preferences over generators based on past satisfaction, cost, latency, or trust 
& Longitudinal utility, retention rate, switching behavior \\

\hline
Generators 
& Maximize user demand under quality--cost tradeoffs 
& Adapt interaction protocols and retrieval configurations to remain competitive 
& User utility, cost efficiency, market share, user retention \\

\hline
Routers 
& Efficient allocation of queries to downstream services 
& Exclusive routing (vertical integration) vs.\ open routing (market-wide discovery) 
& Counterfactual regret, allocation efficiency, diversity, and fairness \\

\hline
Retrievers 
& Remain selected by generators under competition 
& Compete via domain specialization, personalization, or robustness 
& Marginal utility contribution, selection frequency, long-term survival \\

\hline
\end{tabularx}
\label{tab:stakeholders}
\vspace{-10pt}
\end{table*}

To understand how IA agent marketplace should be simulated, we question how we can effectively model users (RQ 1.1) and agents (RQ 1.2). 
To do so, we discuss each stakeholder and their objectives in the marketplace. Table~\ref{tab:stakeholders} summarizes the primary stakeholders in an IA marketplace.
Rather than viewing these stakeholders as static system components, we model them as \emph{strategic actors} that pursue objectives, adapt their behavior under competition, and are evaluated based on their long-run impact on marketplace outcomes.

Across all stakeholder types, we need to consider:
\begin{enumerate}[leftmargin=1.7em]
  \item a goal that defines what success means for the stakeholder,
  \item adaptation strategies that determine how behavior changes over time, and
  \item evaluation signals that reflect whether those strategies succeed in the marketplace.
\end{enumerate}

\subsection{RQ1.1: Modeling Users}

\subsubsection{\textbf{Users: Demand-Side Adaptation.}}
Users represent the demand side of the marketplace.
Their objective is to maximize utility from interactions, accounting for factors such as answer quality \cite{hassan2010beyond}, cost \cite{automix}, and trust \cite{kang_explainitai_2025}.
Referring to the example in Section~\ref{sec:overview}, the expert user may be willing to incur higher cost to access a more capable generator, whereas the general user may prioritize lower cost or faster responses. As a result, each user type operates under a different notion of utility.

Based on their notion of utility, users may adapt by reallocating their demand across competing generators based on past experience \cite{lucherini:trecs, Jiang19degenerate}.
This adaptation may take the form of gradual preference shifts, exploration of alternatives after poor outcomes, or abandonment of previously favored services.

Accordingly, users can be evaluated indirectly, through the signals they generate \cite{hassan2010beyond}.
Longitudinal utility trajectories, retention rates, and switching behavior can serve as market-level feedback that shapes competition among downstream stakeholders.

\subsubsection{\textbf{Simulation of Users}}\label{subsubsec:simualtion_of_users}

Grounded in the stakeholder analysis above, user behavior in the simulation can be modeled through an explicit utility function. For a user $u$ interacting with agent $a$ at time $t$, utility may be defined as
$
\theta_\mu^{u,a}(t) =
\alpha_u \, Q_{u,a}(t)
-
\beta_u \, C_a(t)
-
\gamma_u \, L_a(t)
$,
where $Q$ denotes answer quality, $C$ cost, $L$ latency, and the coefficients capture heterogeneous user sensitivities and encode tradeoff across competing attributes.

Given realized utility, user preferences over agents can be viewed in a perspective of choice modeling across heterogeneous agents \cite{varian_intermediate_2010}.
Different agents can pursue distinct objectives under shared constraints, and their interactions jointly determine marketplace outcomes. 
In choice modeling, decisions are formalized as utility maximization over bundles of commodities \cite{Greene2009}, directly aligning with our formulation.
One simple option, is stochastic choice model \cite{varian_intermediate_2010} that assigns selection probabilities to agents as a smooth function of their estimated utility.

Preference modeling choices are tightly coupled to the construction of the dataset $\mathcal{D}$.
While $\mathcal{D}$ may be instantiated using existing benchmarks or synthetically generated queries \cite{rahmani2025syndl, carterette:simulating-user-behavior, lucherini:trecs, zhang2024usimagent}, the structure and topical distribution of the dataset implicitly determines the space over which user preferences are defined. 
This raises another question: whether user preference should be expressed by topic-dependent distribution, or at the level of persistent users whose preferences evolve across simulation steps.

Prior work has proposed taxonomies of IA behaviors \cite{shelby_taxonomy_2025, chatterji2025howpeopleusechatgpt, filice_generating_2025} which can be derived from interaction logs \cite{zhao_wildchat_2023, joko_wildclaims_2025}. However, it remains unclear how these should be operationalized as sampling distributions or state variables within $\theta_u$ when users face multiple substitutable services. 
Understanding this coupling between dataset design and preference dynamics is central to modeling user adaptation in a marketplace with multiple substitutable services.

Another interesting direction concerns cross market competition formed by users.
Users may substitute across IA stakeholders based on task characteristics.
For example, they may choose a traditional search engine over than a conversational system for certain queries \cite{caramancion2024large}.
Competition therefore occurs not only within a market, but also across markets with partially substitutable functionality.

Beyond structural choices in user simulation, marketplace evaluation invites a deeper examination of behavioral assumptions.
Behavioral economics \cite{NBERw7948} suggests that, within the user population, repeated exposure to multiple substitutable agents may introduce effects that are central to behavioral economics, such as framing \cite{nelson1997toward}, constructed preferences \cite{varian_intermediate_2010}, and network effects \cite{uzzi1996sources}.
Incorporating these phenomena into $\theta_u$ raises questions about which behavioral mechanisms materially affect marketplace outcomes.
For example, popularity signals or exposure biases \cite{exposure_effect} introduced by their selection policies may amplify market concentration.

\subsection{RQ1.2: Modeling Agents}

\subsubsection{\textbf{Generators: Competing for Demand under Constraints.}}

Generators are often supply-side actors interfacing directly with users, especially for GIO creations \cite{swirl_18}. 
From business perspective \cite{jannach2019measuring}, their primary objective is to attract and retain users while operating under quality-cost tradeoffs \cite{kaplan2020scaling} imposed by inference scaling, retrieval, or orchestration choices \cite{dang2025multiagent}.
Looking at the example in Section~\ref{sec:overview}, a more capable generator may prioritize answer quality over traffic, as its higher per-query cost can still yield substantial revenue per user, whereas moderate models may rely on higher traffic volume to compensate for lower revenue per interaction.

Such generators can be modeled as optimizing an expected utility function that balances user-derived utility signals with operational cost.
Adaptation then corresponds to updating retrieval strategies, orchestration depth, or model configuration to maximize long-run utility or market share.

Within the marketplace, generators may operate without retrieval \cite{asai2024selfrag}, rely on a single retriever \cite{lewis:reml-nlg}, aggregate evidence from multiple retrievers \cite{kalra-etal-2025-mor}, or engage in more complex multi-step or agentic retrieval strategies \cite{chu-etal-2024-beamaggr, jin2025searchr1}. 

Evaluation of generators should therefore extend beyond answer quality to include sustained user utility, market share \cite{longpre2025economiesOfOpenIntelligence, openrouter}, cost efficiency \cite{kapoor2025ai}, and retention.

\subsubsection{\textbf{Routers: Market Intermediaries.}}

Routers mediate access to downstream services by allocating queries among competing agents.
In the Section~\ref{sec:overview}'s example, if the generator lacks an effective updating policy, it may depend on the router for retriever selection. However, if routing decisions consistently yield poor outcomes, the generator may bypass it in subsequent interactions.

The two common routers:
user-to-generator routers \cite{hu2024routerbench,ong2025routellm,huang2025lookaheadrouting,agrawal2025llmrankrouting} and generator-to-retriever routers \cite{tang2025adapting,kim2025ltrr,kalra-etal-2025-mor,yeo2025universalrag} differ in position but share a common objective.

A router can be modeled as optimizing an allocation utility that aggregates downstream performance, cost, and fairness constraints.
Preference updates correspond to adjusting routing probabilities based on observed utility or regret signals over time.

Routers may adopt exclusive or open routing policies \cite{openrouter}, shaping exposure and competitive pressure.
Evaluation should therefore focus on allocation-level outcomes such as assignment efficiency \cite{jeong-etal-2024-adaptive}, regret \cite{bubeck2012regret, tang2025adapting}, and exposure diversity or fairness \cite{diaz_ee, kim_fairrag}.

\subsubsection{\textbf{Retrievers: Competing for Downstream Selection.}}

Retrievers supply evidence to generators and compete for repeated selection.
Their objective can be modeled as maximizing marginal contribution \cite{salemi24evaluating} to generator utility while maintaining efficiency and robustness.

Retriever adaptation may manifest as domain specialization \cite{barron2024domain}. Because retrievers are tied to underlying collections, differences in coverage can implicitly correspond to different data providers.
Retrievers may also differentiate through personalization to generators \cite{salemi2024towards}. In this setup, retrievers that fail to provide consistent marginal value \cite{salemi24evaluating} will gradually lose traffic.

Evaluation can therefore focus on marginal contribution, as well as robustness \cite{retrieval_robustness} and efficiency \cite{splade_efficiency,retreival_effec_effi_tradeoff}.

\section{RQ2: Marketplace Metrics}\label{sec:rq2}

In Section \ref{sec:motivation}, together with the experiment (\S\ref{sec:motivating-exp}), we highlighted the limitations of traditional accuracy-based metrics for evaluating agents deployed in competitive environments.
While relevance, correctness, and utility remain foundational, they do not capture how agents perform \emph{as market participants}, where success is reflected in sustained usage, traffic allocation, and competitive positioning.
This motivates the need for \emph{marketplace metrics} that characterize both individual agent outcomes and emergent marketplace dynamics.

Under RQ2, we examine how to evaluate performance in a marketplace setting.
Specifically, we distinguish between metrics that measure the market performance of individual agents (RQ 2.1) and metrics that characterize overall market conditions from the perspective of platform operators (RQ 2.2). 

Accordingly, we propose two complementary classes of metrics.
\textit{Agent-Level Metrics} quantify the position and performance of individual agents within the marketplace, such as their share of traffic and their ability to retain users.
\textit{Marketplace-Level Metrics} capture aggregate structural properties, including concentration and inequality, which reflect dominance, competitiveness, and overall market health.

\subsection{RQ2.1: Agent-Level Metrics}
As shown in Figure~\ref{fig:marketshare-exp}, we derive agent market share from the interaction logs of the simulation.
Rather than computing market share cumulatively over the entire horizon, we use a sliding window to capture temporal dynamics and short-term competitive effects.

Let $A$ denote the population of information access agents in a given market, and let $q_u(t)\in A$ denote the agent selected by user $u\in U$ at timestep $t$.
For a window of length $w$, the market share (MS) of agent $a\in A$ at time $t$ is defined as
\begin{equation}
    \text{MS}_{a}(t; w) = 
    \frac{
        \sum_{u\in U}
        \sum_{k=t-w+1}^{t}
        \mathbf{1}\!\left[q_u(k)=a\right]
    }{
        \sum_{u\in U}
        \sum_{k=t-w+1}^{t} 1
    },
\label{eq:marketshare}
\end{equation}
where $\mathbf{1}[\cdot]$ denotes the indicator function, which equals $1$ when its argument is true and $0$ otherwise.
Therefore, $\text{MS}_{a}(t; w)$ measures the fraction of total interaction volume allocated to agent $a$ within the window.
Tracking $\text{MS}_{a}(t;w)$ over time reveals how agents gain or lose traffic as users adapt their preferences and routing policies evolve.

Beyond traffic volume, an agent's long-term viability depends on its ability to retain users after initial adoption.
Inspired by customer retention metrics from the business literature \cite{keiningham2007value}, we define customer retention (CR) measures tailored to agent marketplaces.

Let $\tau_{u,a} = \min \{\, t \in \{1,\dots,T\} : q_u(t)=a \,\}$ denote the first timestep at which user $u$ selects agent $a$, if such a timestep exists.
Given a window length $m$, the \emph{user-level retention} of agent $a$ for user $u$ is
\begin{equation}
    \text{CR}_{u,a}(m) = 
    \frac{1}{m}\sum_{t=1}^{m}
    \mathbf{1}\!\left[q_u(\tau_{u,a}+t)=a\right],
\label{eq:retention-user}
\end{equation}
which measures the fraction of the next $m$ interactions, following first adoption, in which $u$ continues selecting $a$.
Aggregating across users yields the \emph{agent-level retention}
\begin{equation}
    \text{CR}_{a}(m) = 
    \frac{1}{|U_a|}
    \sum_{u\in U_a}\text{CR}_{u,a}(m),
\label{eq:retention-agent}
\end{equation}
where $U_a := \{ u \in U : \tau_{u,a}\ \text{is defined} \}$ is the set of users who have tried agent $a$ at least once.
A higher $\text{CR}_{a}(m)$ indicates that users who initially sample $a$ tend to continue allocating a larger fraction of their subsequent interactions to that agent.

However, higher market share does not necessarily imply higher customer retention. A model may maintain a small market share yet achieve strong retention by consistently delivering high quality to a targeted user segment, as in Section~\ref{sec:overview}, where a more capable model provides exclusive value to the domain expert.

\subsection{RQ2.2: Marketplace-Level Metrics}
Beyond agent-level metrics, marketplace-level metrics characterize aggregate interaction patterns and diagnose higher-level market properties such as concentration, convergence, and inequality. These metrics can be computed from the cumulative interaction behavior observed in the simulation logs.
These marketplace-level metrics primarily reflect the interests of mediation platform owners or routers rather than individual agents. If the market becomes highly concentrated, routing may become trivial or even unnecessary, and competing service providers may progressively lose their share of demand.
Here, standard market concentration and inequality metrics from the economics literature can be applied \cite{longpre2025economiesOfOpenIntelligence}.

The Herfindahl--Hirschman Index (HHI) \cite{rhoades1993herfindahl} captures the degree of market concentration.
Using the windowed market-share from Equation \eqref{eq:marketshare}, we define HHI-based market concentration as
\begin{equation}
    \text{HHI}(t;w) = 
    \sum_{a\in A} \text{MS}_{a}(t;w)^2.
\label{eq:hhi}
\end{equation}
Larger $\text{HHI}(t;w)$ indicates greater concentration, approaching monopoly as the value increase.

Marketplace outcomes may also be evaluated through fairness lenses, e.g., whether providers receive exposure consistent with a target allocation.
Notion of \textit{fair} market share helps us creating related marketplace-level metrics. Let's define a \textit{target} share or exposure distribution $\epsilon^*$ over agents, where $\sum_{a\in A}\epsilon^*_a=1$. 
Depending on the definition of fairness, we can define $\epsilon^*_a$ as $1/|A|$ for equal share, or proportional to the merit of an agent, where the merit can be based on the static benchmark score (as in Table \ref{tab:static_results}) or cumulated $\mu$'s during the window $w$ at time $t$.

From here, we can yield a complementary market dominance measure
\begin{equation}
    \Delta(t;w) = 
    \max_{a\in A}\text{MS}_{a}(t;w) - \epsilon^*_a.
\label{eq:dominance-gap}
\end{equation}
A larger $\Delta(t;w)$ indicates that the top agent captures substantially more market share than the target share.

Additionally, inspired by expected exposure (EE) in document ranking \cite{diaz_ee}, we can define an expected exposure difference metric by comparing realized traffic shares to a target exposure distribution $\epsilon^*$ over agents.
Using market share as a proxy for exposure, we define expected exposure difference as 
\begin{equation}
    \text{EE}_{\text{marketplace}}(t;w) = 
    \|\text{MS}_{a}(t;w) - \epsilon^*_a\|_2^2 ,
\label{eq:eed-ms} 
\end{equation}
where $\|\cdot\|_2^2$ denotes the squared $\ell_2$ norm.

Following the definition of expected exposure disparity (EE-D) \cite{diaz_ee}, we obtain $\text{EE-D}_{\text{marketplace}}(t;w) = \|\text{MS}_{a}(t;w)\|_2^2$,
which is analogous to the $\text{HHI}(t;w)$.

\section{RQ3: Adoption to Evaluation Campaigns}\label{sec:rq3}
The marketplace evaluation framework also raises questions about how large-scale evaluation campaigns, such as TREC \cite{soboroff2023overview_trec32} and CLEF \cite{clef}, can evolve beyond traditional Cranfield-style setups. 
Historically, these campaigns have relied on a static test collection and offline run submission, where systems are evaluated independently by organizers or assessors. While this paradigm has enabled reproducibility and straightforward comparability, it does not capture the dynamics we want to measure.

Adopting a simulated marketplace environment within evaluation campaigns introduces new design choices that can be organized along two orthogonal axes.

\textbf{Axis 1: Peer Competition Vs. Benchmark Competition.}
In a peer competition setting, submitted systems compete directly with one another for traffic and exposure within a shared marketplace simulation. Outcomes will be jointly determined by the participant pool, enabling analysis of dominance, switching, and concentration under head to head conditions. However, results may depend on the specific mix of systems submitted in a given year, potentially limiting stability and cross-year comparability. 
In a benchmark competition setting, each submission competes against a fixed population of organizer-provided IA agents. This improves reproducibility, allowing clearer attribution of outcomes to individual system properties.

\textbf{Axis 2: Run Submission Vs. Agent Submission.}
In the run submission model, participants provide static outputs that are treated as cached agent responses, and only the simulated user population adapts over time. This preserves many practical advantages of traditional campaigns, including low computational burden, and reproducibility. However, systems themselves cannot adapt or respond to competitive feedback. In the agent submission model, participants submit executable agents that interact with a shared simulation, making sequential decisions under competition. This enables evaluation of adaptive routing, cost-aware strategies, and strategic behavior, but introduces challenges in controlling stochasticity, ensuring fair resource allocation, and maintaining reproducibility.

\textbf{Task Design.}
From a task-design perspective, marketplace-based campaigns could focus on settings where interaction and adaptation are essential, such as distributed information retrieval (e.g., TREC Million LLMs \cite{kanoulas2025millionLLM}), generator–retriever coordination (e.g., TREC RAG \cite{trec-rag}), and interactive information access (e.g., TREC iKAT \cite{trec_ikat}). These tasks would allow campaigns to evaluate not only effectiveness, but also robustness, adaptability, and strategic behavior under competition.

\section{Validation of Marketplace Simulation}
\label{sec:metaeval}
Validation of marketplace simulations is a necessary step following any simulated evaluation \cite{allen2001introduction, rand_agent-based_2011}. 
Because simulation outcomes emerge from interacting components rather than isolated system runs, validation must assess whether the constructed environment meaningfully reflects the phenomena it aims to study.

We organize validation around three pillars.  Validation of (1) sampled or synthetically generated user queries; (2) marketplace interaction dynamics and stakeholder modeling; and (3) marketplace metrics (meta-evaluation).

\textbf{Validation of User Queries.}
Validation of user queries concerns whether sampled or synthetically generated queries adequately represent the intended demand distribution. 
From a construct validity perspective, the query distribution should align with the real user population it claims to model. 
This can be examined through distributional comparisons against real logs when available \cite{joko_wildclaims_2025, shelby_taxonomy_2025}, or through expert review of coverage across task types, difficulty levels, and user expertise strata.

Convergent validity can be assessed by measuring system rank correlations between simulation-based evaluations and established offline benchmarks. 
Suppose we obtain systems from a public model leaderboard that reports download counts or market share. We can instantiate those systems within our marketplace simulation and run the simulation under controlled conditions. Based on a chosen marketplace metric, we can produce a ranking of systems from the simulation. We then compute the correlation between the simulated ranking and the ranking implied by the real-world leaderboard. A higher correlation would suggest that the simulation captures relevant aspects of real-world competitive dynamics.

\textbf{Validation of Marketplace Stakeholders.}
As another example of construct validity, we can introduce controlled perturbations within the marketplace setup and examine whether the intended marketplace-level or agent-level metrics respond accordingly. For instance, if we artificially increase the latency of a particular model, we would expect to observe lower customer retention or increased market concentration due to the system's degradation.

\textbf{Meta Evaluation of Marketplace Metrics.}
Marketplace metrics require meta evaluation to ensure they meaningfully support comparison \cite{sakai2006evaluating}. 
Key properties include sensitivity, discrimination power, and robustness to gaming. 
Sensitivity can be assessed through bootstrap based hypothesis testing and ASL curves. 
Discrimination power measures a metric's ability to separate agents with genuinely different performance profiles (e.g., if the current definition of customer retention rate has a strong discrimination power is unclear). 
Robustness to gaming evaluates whether agents can inflate scores without improving underlying utility. 
Grounded in measurement theory, these analyses strengthen the credibility of marketplace simulation as an evaluation framework.

\section{Further Research Agenda}
\subsection{Single-Market Vs. Multi-Market}
As shown in Section~\S\ref{sec:motivating-exp}, evaluation can be conducted using only user-facing agents without explicitly modeling downstream interactions. Recall a \emph{market} denotes a task-specific set of stakeholders that jointly deliver a functionality and compete for traffic within it. We refer to such settings as \emph{single-market} evaluations. Here, routers that mediate user-to-generator selection remain within the same market. Although front-facing agents may rely on complex internal components, evaluation considers only their observable end outcomes (e.g., final responses or rankings), abstracting away internal or downstream dynamics.

Single-market evaluation can be naturally extended to \emph{multi-market} evaluation by explicitly modeling multiple coupled markets within the same simulation.
As exemplified by RAG, such settings involve distinct markets that are linked through a compositional workflow (Figure~\ref{fig:simulation-snapshot}).

In contrast to single-market evaluation, multi-market evaluation explicitly models inter-market coupling and captures how competition, mediation, and composition jointly shape user-facing outcomes. For example, a generator's realized quality depends on the retriever it invokes, while a retriever's value depends on how generators use its evidence. By treating routers and retrievers as first-class stakeholders and logging full compositional trajectories, including which agents were selected in each market and with what outcomes, multi-market evaluation supports cross-market diagnosis and enables attribution of observed behaviors to specific markets or mediation policies.

\subsection{Economic Models of Competition}
Broadly, the marketplace perspective motivates closer engagement with economic models of competition \cite{varian_intermediate_2010}.
Classical concepts such as market dominance \cite{harrington_co-evolution_2005} and competitive entry \cite{polo2018entry} have well-established theoretical foundations, but their application to AI and information access agent marketplaces remains largely unexplored. Also, as discussed in Section~\ref{subsubsec:simualtion_of_users}, incorporating choice modeling for agent selection and behaviorally grounded user and agent models~\cite{NBERw7948} remains underexplored in information retrieval.
Such perspectives can enrich the study of retrieval in \textit{multifirm} settings.

Beyond being helpful to information retrieval research, this framework may also benefit scholars in computational economics and algorithmic market design.
By providing a controllable simulation environment with explicit user adaptation, agent competition, and market-level observables such as market share, retention, and concentration, it offers a testbed for studying dynamic competition among AI agents under realistic demand feedback.
Such a setting enables the empirical exploration of classical economic constructs, including dominance, entry, switching costs, and equilibrium formation, in domains where analytical modeling alone may be insufficient.
In this sense, marketplace simulation and evaluation framework provides a new computational laboratory for economic research on digital competition.

\subsection{Optimization of Agents and Adversaries}
Marketplace evaluation naturally raises the question of how information access agents should be optimized when success is defined by competitive, market-level outcomes along with effectiveness metrics.
Simulated environments have played an important role in optimizing agents in interactive settings \cite{peng-etal-2018-deep}, particularly in reinforcement learning, where feedback is delayed and behavior unfolds over time.

In a marketplace, optimization can be framed as an online decision problem where agents repeatedly choose actions (e.g., agent selection \cite{kanoulas2025millionLLM}) under uncertainty about competitor behavior.
This setting naturally supports the study of bandit \cite{tang2025adapting} and online learning-to-rank methods \cite{dai2017ltrResources, zoghi2017online} for agent selection that optimize long-term objectives such as retention, or profit under cost constraints.

The optimization of agents in a marketplace naturally brings attention to adversarial behaviors.
Agents may deliberately or unintentionally exploit weaknesses in evaluation metrics, routing policies, or user models, leading to outcomes that are harmful to the marketplace as a whole.
This includes behaviors analogous to reward hacking \cite{ibarz2018reward}, where agents optimize for observed signals while undermining the intended objectives of the system.
Understanding such behaviors raises questions about the robustness of marketplace metrics, the vulnerability of routing mechanisms, and the safeguards required to ensure healthy competition \cite{olteanu2021facts}.

\subsection{Implementation Choices}
Marketplace simulations require several practical design decisions that can affect outcomes.
Key choices include the number of users, and the number of competing systems. The batch size of users to sample per simulation step may influences the effective time scale of adaptation. The simulation horizon determines whether results reflect short term dynamics or long term convergence. Finally, from sampled users, simulations may operate synchronously, with global in-batch updates, or asynchronously, with staggered interactions that more closely resemble real deployment.

\subsubsection{Resource Contribution}
To facilitate systematic exploration of these factors, in the interest of feasibility and reproducibility, we release a Python package \texttt{marketplace-eval} that implements the proposed framework. The package provides a minimal starting point with example configurations, including the exact simulation setup used in our motivating experiment, enabling researchers to readily replicate results and extend to new settings. Beyond parameter control over population sizes, batch sampling, horizon length, and interaction mode, the package supports modular overrides of core components, allowing researchers to customize interaction dynamics and adapt the framework to a wide range of experimental settings.
The code is available at \url{https://github.com/kimdanny/marketplace-eval}.

\section{Conclusion}
\label{sec:conclusion}
We propose a marketplace-based perspective for evaluating information access agents, shifting evaluation from isolated system performance to outcomes shaped by interaction, adaptation, and competition.
We outline a research agenda and release extensible resources to support further study of marketplace evaluation.
This perspective encourages evaluation frameworks that better reflect real-world deployment, where information access systems co-evolve with users and other agents.

\begin{acks}
This work was supported in part by the Center for Intelligent Information Retrieval, and in part by the National Science Foundation under Grant No. 2402873 and 2402874.
Any opinions, findings and conclusions or recommendations expressed in this material are those of the authors and do not necessarily reflect those of the sponsors.
\end{acks}

\bibliographystyle{ACM-Reference-Format}
\bibliography{XX-references.bib}

\end{document}